\pdfoutput=1

\documentclass{PoS}
\usepackage{amsmath}

\title{Exploring a hidden center symmetry with electrically charged quarks}

\ShortTitle{Exploring a hidden center symmetry with electrically charged quarks}

\author{\speaker{Sam R. Edwards}\\
       Institut f\"ur Kernphysik, Technische Universit\"at Darmstadt, D-64289 Darmstadt, Germany\\
        E-mail: \email{edwards@crunch.ikp.physik.tu-darmstadt.de}}

\author{Andr\'e Sternbeck\\
        Institut f\"ur Theoretische Physik, Universit\"at Regensburg, D-93040 Regensburg, Germany\\
        E-mail: \email{andre.sternbeck@physik.uni-regensburg.de}}

\author{Lorenz von Smekal\\
        Institut f\"ur Kernphysik, Technische Universit\"at Darmstadt, D-64289 Darmstadt, Germany\\
        E-mail: \email{lorenz.smekal@physik.tu-darmstadt.de}}

\abstract{
It is usual to study confinement via quantum chromodynamics (QCD) alone. The deconfinement transition of the pure gauge theory (i.e. with static quarks) is then characterized by the breaking of center symmetry. Center vortices offer an intuitive and quantitative description of the transition. Dynamical quarks explicitly break center symmetry, and the phase transition becomes a crossover. However, it may be misleading to study QCD in isolation. Quarks also carry fractional electric charge.  This bestows the Standard Model with a global center symmetry that combines color center phases with an appropriate electromagnetic phase.  Is this symmetry relevant to confinement?  We begin our investigation by studying a 2-color model of QCD with half-integer electrically charged quarks. 

}

\FullConference{The XXVIII International Symposium on Lattice Field Theory\\
		 June 14-19,2010\\
		 Villasimius, Sardinia Italy}

\begin{document}

\section{Introduction}
The finite temperature deconfinement transition of pure SU($N$) gauge theory is intimately tied to center symmetry. The expectation value of the Polyakov loop $\langle P \rangle$ jumps from zero to a finite value at $T_c$, thereby picking a center sector and signaling deconfinement along with the spontaneous breaking of center symmetry. The transition is well described by the dynamics of spacelike center vortices, whose proliferation at low temperatures disorders the Polyakov loop and leads to confinement \cite{Greensite:2003bk}. They carry center flux and separate regions where the Polyakov loop differs by a phase $z\in Z_N$. The suppression of center vortices at high temperature coincides with the ordering of the Polyakov loop, and their free energy offers an elegant dual order parameter for the transition \cite{DeForcrand:2001dp, DeForcrand:2002}. Deconfinement mirrors the order-disorder transition of a simple $Z_N$ spin model  \cite{Svetitsky:1982gs}, with vortices playing the role of spin interfaces. When both phase transitions are of second order their behavior is universal. We have fruitfully exploited this for SU(2) and SU(3) in 2+1 dimensions, where many exact results are available from the corresponding spin models \cite{Edwards:2009qw,Lat2010}.

Center symmetry is explicitly broken by the introduction of quarks, however, and this picture is lost. Their ordering effect is easily seen via a hopping expansion of the fermion determinant. One obtains Polyakov loop terms that favour the $\langle P \rangle = 1$ center sector in the same way that an external magnetic field aligns spins in a spin system. 

Dynamical quarks also make the role of center vortices unclear. 
Vortex sheets, which were freely moved in the pure gauge ensemble, become observable. Quarks pick up the center flux as a phase when they encircle a vortex. But what about the quarks' fractional electric charge? Phases from the center of the color gauge group may be compensated by an appropriate electromagnetic contribution. Phase cancellation renders a combined center vortex and electromagnetic Dirac string unobservable. This reflects a hidden global gauge symmetry of the Standard Model. It combines the centers of each group factor and is therefore an extension of center symmetry. 

We would like to study the relevance of this symmetry to confinement and the phase structure of the Standard Model. As a first step we consider 2-color QCD with the addition of quarks carrying fractional electric charge. Here we concentrate on the possibility that the quarks' electromagnetic coupling leads to disorder in the color fields, akin to placing a spin system in a random external field.

\section{The hidden symmetry}
When subjected to a nontrivial SU(3) center gauge transformation, or upon encircling a center vortex, quarks pick up a phase $e^{\pm i2\pi/3} \in Z_3$. Given their electric charges, $Q=\frac{2}{3}e$ or $-\frac{1}{3}e$, an additional electromagnetic factor $e^{\pm i2\pi Q/e}$ is enough to exactly cancel the color contribution.  This is in fact the phase that a particle with electric charge $Q$ receives when it encircles the Dirac string of a monopole with minimal magnetic charge with respect to $e$. Colorless particles with integer electric charge are blind to both phases by default. Therefore, the pair of combined transformations \begin{equation}
(e^{i2\pi/3},e^{i2\pi Q/e}),\;(e^{-i2\pi/3},e^{-i2\pi Q/e})\in \text{SU(3)}\times\text{U(1)}_{em} 
\end{equation}
act trivially on all particles in the Standard Model. Together with the identity transformation they form a global $Z_3$ symmetry. The symmetry doubles when we include the weak interactions.
Electric charge $Q$ is related to hypercharge $Y$ and the third component of weak isospin $t_3$ by $Q/e=t_3+Y/2$.  Since $e^{i2\pi t_3}$ represents $-1\in$ SU(2), the electromagnetic phase $e^{i2\pi Q/e}$ is given by $(-1,e^{i\pi Y})\in$ SU(2)$\times$U(1)$_Y$. The discrete symmetry group is therefore generated by 
\begin{equation}
(e^{i2\pi /3},-1,e^{i\pi Y})\in \text{SU(3)}\times \text{SU(2)}\times \text{U(1)}_Y. 
\end{equation}
This gives six elements, so the 'true' or minimal gauge group of the Standard Model is \\ SU(3)$\times$SU(2)$\times$U(1)$_Y/Z_6$. See \cite{Baez:2009dj} for a review. 


This $Z_6$ symmetry is very important in the context of grand unified theories. Without it, the Standard Model gauge group would be too large to fit in SU(5) $\subset$ SO(10) . Moreover, it endows the model with non-trivial topology that allows for defects carrying both color and electromagnetic flux. These are, namely, combined center vortex and electromagnetic Dirac strings/surfaces.
Given the importance of center vortices in pure SU($N$) gauge theory, it is natural to ask if the discrete symmetry and combined vortices are relevant to the confinement of dynamical quarks.\footnote{The $Z_6$ symmetry was recently studied on the lattice by Bakker, Veselov and Zubkov \cite{Bakker:2005ph}. There they were more interested, however, in consequences for electroweak and Higgs physics.}
This type of defect unification has already been shown to have a large impact on the phase diagram of a toy U(1)$\times$U(1) model \cite{vonSmekal:2005mq,vonSmekal:2006zx}. 

\section{Toy model}
Lattice simulations of the full Standard Model gauge group with dynamical quarks are beyond our reach. As such, we start with a toy model of 2-color QCD plus electromagnetism with 2 flavors of Wilson fermions in 3+1 dimensions. Our 'up' and 'down' quarks have fractional charge $\pm \frac{1}{2}e$ relative to the U(1)$_{em}$ gauge action. The weak interactions are discarded. Nevertheless, we have a chance to see what interesting physics may come from the quarks' non-trivial coupling to both photons and gluons. The lattice action is 
\begin{equation}
S = -\sum_{\text{plaq.}}\left( \frac{\beta _{col}}{2}\text{Re Tr }U_p +\beta _{em} \cos \theta_p\right) +  S_{f,W},
\end{equation}
where $S_{f,W}$ is the usual Wilson fermion action with the distinction that the parallel transporters for quarks are  products of an SU(2) color matrix and a U(1)$_{em}$ phase. They are of the form,
\begin{equation}
\label{eqn:transporters}
 U_\mu(x) e^{i\theta_{\mu}(x)/2},\;\; U_\mu (x)\in\text{SU(2)},\;\theta_{\mu}(x)\in (-2\pi,2\pi].
\end{equation}
The SU(2) plaquettes $U_p$ and U(1) plaquette angles $\theta_p$ are formed from $U_\mu$ and $\theta_\mu$ in the usual way.
The fractional charge of the quarks is encoded in the fact that their parallel transporters contain \emph{half} the U(1) angle relative to the $\theta_\mu$ that appear in the plaquette angle $\theta_p$. For instance, what the quarks see as an $e^{i\theta_\mu/2}=-1$ electromagnetic link appears as a $e^{i\theta_\mu}=+1$ link in the gauge action. Since the 'size' of the compact U(1) is determined by the quarks' electric charge, the range of $\theta_{\mu}(x)$ is chosen such that we integrate over all possible electromagnetic transporters for the quarks.  

The gauge fields are blind to global SU(2) center  and U(1) transformations, so we would have a global $Z_2\times$U(1) center symmetry in the absence of fundamental fields. It is clear from Eq. \eqref{eqn:transporters} that a combined color center phase $-1\in$ SU(2) and electromagnetic phase of $-1$ acts identically on the quarks. Thus a $Z_2$ subgroup of this symmetry survives their introduction, and the minimal gauge symmetry of our model is  SU(2)$\times$U(1)$_{em}$/$Z_2$. $Z_2$ plays the role of the hidden symmetry of the Standard Model.

The loops in a hopping expansion of the fermion determinant are now constructed from both color and electromagnetic parallel transporters. For $N_t=4$ time slices, the leading contribution comes from plaquette and Polyakov loop terms. Arranged in an effective action, they give

\begin{equation}
\label{eqn:hopping}
S_{f,\text{eff}} = -16\kappa^4 \left( \sum_{\text{plaq.}} \cos{\frac{\theta_p}{2}}\cdot \text{Re Tr }U_p  + 8\sum_{\vec{x}} \cos{P_{\theta/2}} \cdot {\text{Re }{P_{col}}  } \right) +\dots,
\end{equation}
where $\kappa$ is the hopping parameter.
Note that the U(1) contribution to the plaquette term is formed from U(1) link angles as seen by the quarks, i.e. $e^{i\theta_\mu /2}$.  $P_{col}$ denotes the traced SU(2) Polyakov loop, while $P_{\theta/2}$ denotes the U(1) Polyakov loop angle, $P_{\theta/2}(\vec{x})=\sum_{t=0}^{N_t-1}\theta_t(t,\vec{x})/2$. 

In the absence of electromagnetism, the Polyakov loop term has the same effect as coupling an external ordering magnetic field to a spin system. It picks an orientation. The U(1) coupling alters this picture, however, since disorder in the quarks' U(1) links promotes disorder in the SU(2) links.
If $\cos{P_{\theta/2}}$ is disordered, then the Polyakov loop term from the hopping expansion will be minimized only if $P_{col}$ is also disordered. The effect on $P_{col}$ may be compared to placing a spin system in a random magnetic field. For instance, consider an Ising model with the  Hamiltonian 
\begin{equation}
\mathcal{H}=-J \sum_{\langle i,j \rangle}{ {s_i s_j} - h \sum _{i} h _i s_i},
\end{equation}
where $s_i=\pm1$ are spins and the direction of the magnetic field $h_i=\pm 1$ is site dependent and random. Such models have very interesting phase structures. Order-disorder phase transitions are still possible, but their existence and type depend on the magnitude of the applied random field \cite{spinglasses}. 

More realistically, imagine an Ising model in a \emph{fluctuating} external field. If the fluctuations are large they will hinder the ordering of spins, thus preserving the global $Z_2$ symmetry. This is reminiscent of the Peccei-Quinn mechanism in QCD \cite{Peccei:1977}, which replaces the $\bar{\theta}$ parameter of the CP violating term by an axion field. Dynamical preservation of CP symmetry is then achieved if the fluctuations of the axion field average to zero. 

In the case of spins, we can model the interaction with a dynamical field by including additional U(1) variables at each site. This gives a so-called XY-Ising model \cite{Nightingale:1995}. Indeed, the typical coupling of U(1) and Ising energy densities in these systems is similar to the mixed plaquette term from Eq. \eqref{eqn:hopping}. As for spins in a static random magnetic field, the coupling in an  XY-Ising model leads to very novel critical behavior. We may expect the same for our gauge theory. 

Compact QED has a disordered, confining phase for $\beta_{em} \lesssim 1$. Here  the U(1) should have a large effect. However, this is not a physically relevant regime since it confines all electric charges. We would like to confine quarks but not integer charged particles such as electrons. The interesting question is whether the U(1) has an influence on the color dynamics in the Coulomb phase for integer charges, in which our U(1) links are ordered with respect to the pure gauge action. Now the fractional electric charge of  quarks becomes important. The phases $\theta_\mu$ and $\theta_\mu+2\pi$ are distinguished by the quarks, but not by the gauge action. As we cross $\beta_{em}\approx 1$, the U(1) links should become ordered for integer charges, but there is still room for $Z_2$ disorder as seen by quarks. Does this lead to disorder in the color links?

\newpage
\subsection{Dynamical simulations}

\begin{figure}
	\centering
		\includegraphics[width=1\columnwidth]{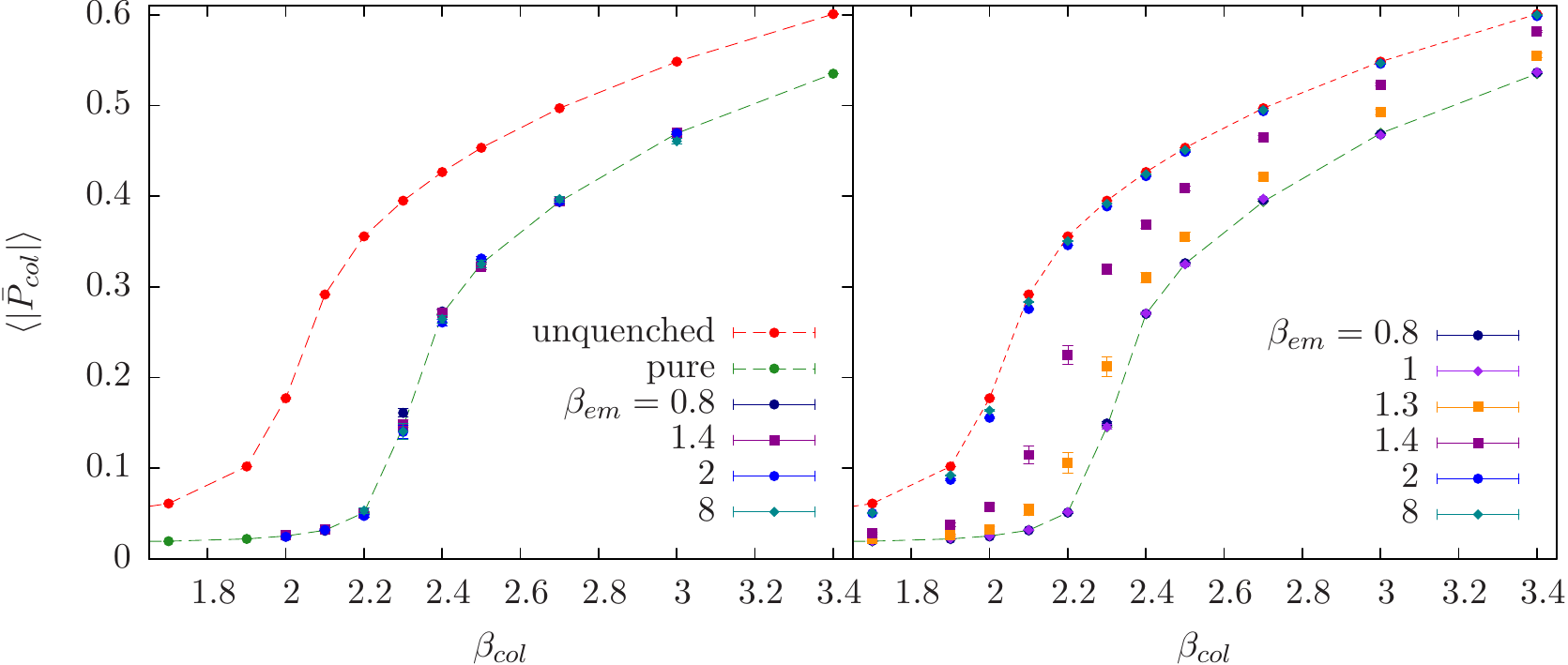}
	\caption{Volume averaged SU(2) Polyakov loop in the presence of quarks with fractional electric charge. On an $8^3\times4$ lattice with $\kappa = 0.15$. The 'unquenched' and 'pure' results are without electromagnetism. The left plot is from a hot start, the right from a cold start. Results with $\kappa = 0.175$ (not shown) are similar.}
	\label{fig:plot-su2polyhot}
\end{figure}

We checked this by measuring the magnitudes of volume averaged Polyakov loops in our SU(2)$\times$U(1)$/Z_2$ toy model with dynamical quarks. We update via Hybrid Monte Carlo.\footnote {From an algorithmic point of view we use 2 degenerate (i.e. equal charge) species of quarks. However, owing to the pseudoreality of the fundamental representation of SU(2), this is equivalent to a system with $\pm \frac{1}{2}e$ quarks. You can convince yourself of this by playing with the charge conjugation operator and color basis.}
For initital testing, we restricted ourselves to  $8^3\times4$ lattices. We have yet to  calculate the mass scale and critical  $\kappa_c$ in our model, so we use the 2-color results of Skullerud et al. for guidance \cite{Skullerud:2003yc}. At zero temperature they find a lattice spacing of $a^{-1}\approx 690$ MeV for $\beta_{col}=1.8$, $\kappa = 0.178$.  From this we chose $\kappa=0.15,\; 0.175$ for 'heavy' and 'light' simulations, respectively.  In Fig. \ref{fig:plot-su2polyhot} we plot the results for the SU(2) Polyakov loop measured near the pure SU(2) phase transition. Various U(1) couplings were used, from the U(1) confined phase ($\beta_{em}=0.8$) to deep into the Coulomb phase for integer charges ($\beta_{em}=8$). We verified that the U(1) Polyakov loop for integer charges (i.e. constructed from $e^{i\theta_t}$ links) jumps from zero to one in the vicinity of $\beta_{em} = 1$, which signals that the compact QED phase transition remains essentially unchanged. 

Using a random, hot start for both the color and U(1) fields,  the pure gauge result is reproduced for the SU(2) Polyakov loop, despite the fact that our model contains dynamical quarks (see Fig. \ref{fig:plot-su2polyhot} left). This is the case for both our 'heavy' and 'light' simulations. We find the same result using a cold start with respect to the U(1) gauge action but with $Z_2$ disorder in the U(1) links as seen by quarks, i.e. $U_\mu=1$, $e^{i\theta_\mu /2}=\pm  1$.  The U(1) gauge action is unable to dynamically remove the $Z_2$ disorder and it becomes frozen in. This disorders the SU(2) links via the coupling to quarks and one recovers the result for the pure gauge theory.

Things are more complicated when we simulate from a completely ordered start ($U_\mu=1$, $e^{i\theta_\mu/2}=1$). The quenched SU(2) behavior persists into the Coulomb phase for integer electric charges, but not indefinitely. We observe a transition to the unquenched SU(2) result near $\beta_{em}=1.3$ (see Fig. \ref{fig:plot-su2polyhot} right). Fluctuations of the link angles are suppressed so strongly that the quarks are unable to generate $Z_2$ disorder.  In this case, the order from our cold start is frozen in and effectively breaks the $Z_2$ symmetry.

\subsection{Hopping expansion simulations}

\begin{figure}
	\centering
		\includegraphics[width=1\columnwidth]{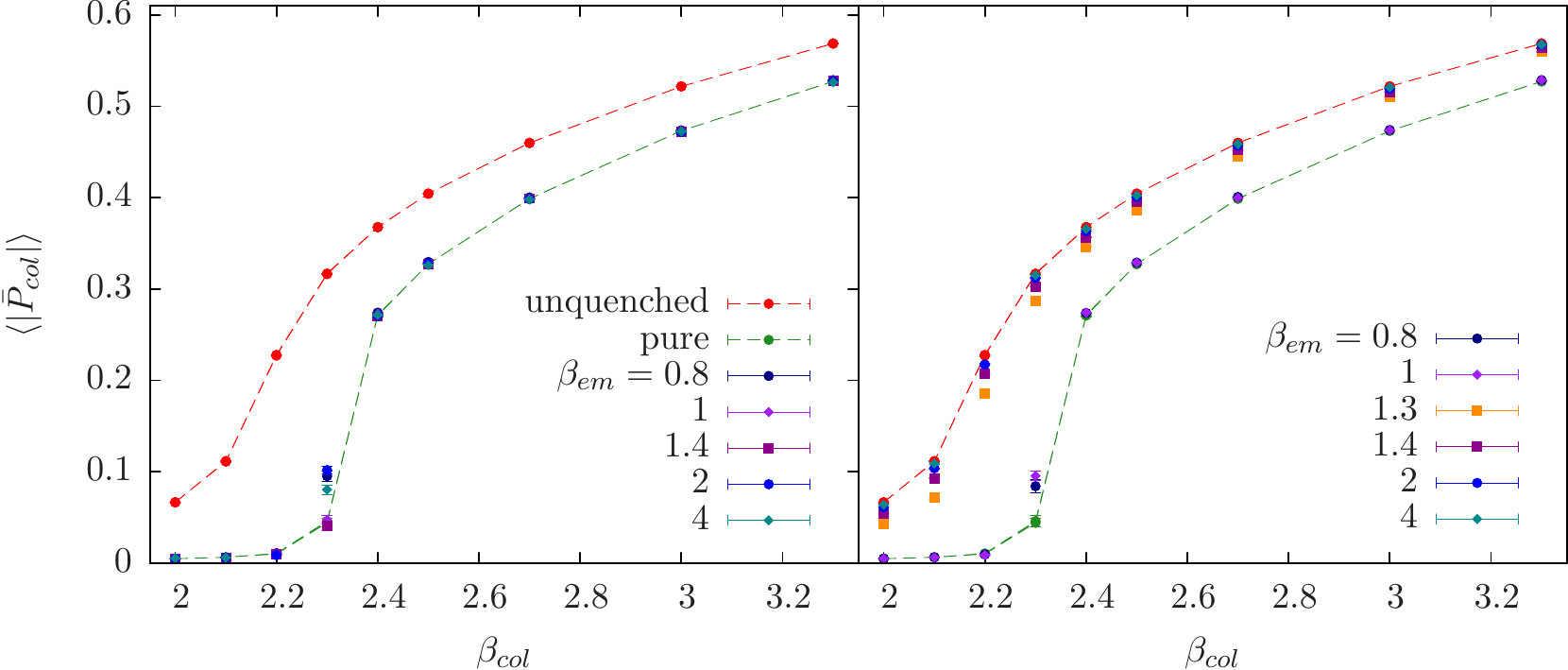}
	\caption{Volume average SU(2) Polyakov loop, where the effect of electrically charged quarks is included up to $\mathcal{O}(\kappa^4)$ in the hopping expansion. On an $24^3\times4$ lattice with $\kappa = 0.175$. The 'unquenched' results are without electromagnetism. The left plot is from a hot start, the right from a cold start.}
	\label{fig:plot-su2polyhothop}
\end{figure}

Before performing full dynamical simulations on larger lattices, we would like to see how far our intuition from the hopping expansion can take us. As such we have run simulations for $N_t=4$ where the effect of quarks are included up to $\mathcal{O}(\kappa^4)$. This means that we include the terms in Eq. \eqref{eqn:hopping} in addition to the pure gauge actions for SU(2) and U(1). We simulate using  $\kappa=0.175$ on $24^3\times 4$ lattices, once more from both hot and cold starts.  The results are shown in Fig. \ref{fig:plot-su2polyhothop}. As for the fully dynamical case, $Z_2$ disorder from a hot start leads to quenched behavior for the SU(2) Polyakov loop. The red points are from simulations of our hopping expansion effective action for SU(2) alone. From a cold start we observe a transition towards this curve near $\beta_{em}>1.3$. We have crosschecked this behavior using both HMC and a mixture of SU(2) heatbath plus U(1) Metropolis updates. This suggests that the retention of $Z_2$ disorder from a hot start is a feature of the model and is not an artifact of our HMC algorithm for fully dynamical quarks.

Since the hopping expansion captures the essential features of the model, it can be exploited to better understand how the fractional electric charge of quarks influences the color dynamics. For instance, the Polyakov loop term has the dominant effect for unquenched SU(2) without electromagnetism. The plaquette term merely shifts $\beta_{col}$. With the coupling to U(1), however, the plaquette contribution from the hopping expansion becomes relevant. It also conveys disorder between the color and electromagnetic links. By simulating with and without each term we can determine their relative importance. This may help us to construct an effective spin model. Via the hopping expansion we can also experiment with large values of $\kappa$. From a hot start one may then expect the color and U(1) links to align with respect to each other but not with respect to the gauge action. These points will be addressed in the future.

\newpage

\vspace*{-1cm}

\section{Discussion}

\vspace{-.2cm}

The results for our SU(2)$\times$U(1)/$Z_2$ model demonstrate that the fractional electric charge of quarks relative to a compact U(1)$_{em}$ gauge action may have a non-perturbative effect on the color sector. It is related to a discrete global symmetry that combines electromagnetic and color phase factors. Although the electromagnetic phases are not distinguished by the U(1) gauge action, disorder amongst them leads back to the color degrees of freedom via the coupling to quarks. This is enough to recover quenched behavior for the SU(2) Polyakov loop, even far into the Coulomb phase for integer charges.  

Since the Standard Model contains such a global symmetry, it may be misleading to study the group factors SU(3)$_{col}$ and U(1)$_{em}$ separately. It is tempting to speculate about how the fractional charge of quarks may alter the 'Colombia plot' picture of confinement \cite{Brown:1990ev}. For QCD with zero chemical potential, a first order transition persists for heavy quarks. If the nontrivial coupling to U(1)$_{em}$ is able to disorder the color fields,  this may extend the first order region to lighter quarks and perhaps sharpen the crossover near the physical quark masses.
Note, however, that our results depend on the initial conditions of the simulations. In order for the U(1)$_{em}$ to have an influence, we need a scenario in which disorder with respect to the electromagnetic phases seen by quarks is frozen in when we reach the perturbative regime for integer electric charges. This leads naturally to a consideration of grand unified theories. It also prompts us to consider the type of field configurations that provide the disorder. We expect these to be vortices that carry both color and electromagnetic flux. Generalizing 't Hooft's twisted boundary conditions \cite{'tHooft:1979uj} to our SU(2)$\times$U(1)/$Z_2$ model, we will be able to study combined vortices in  prescence of dynamical quarks.

\bigskip
\noindent\textbf{Acknowledgements:} This work was supported by the
Helmholtz International Center for FAIR within the LOEWE program of
the State of Hesse, the Helmholtz Association Grant VH-NG-332, and the
European Commission, FP7-PEOPLE-2009-RG No.~249203. Simulations were
performed on the high-performance computing facilities of eResearch
SA, South Australia.

\end{document}